\documentclass[11pt, a4paper, reprint, amsmath,amssymb, aps, pra]{revtex4-2}

\usepackage{graphicx}
\usepackage{dcolumn}
\usepackage{bm}

\usepackage[caption=false]{subfig}

\usepackage{geometry}
\DeclareMathOperator{\sinc}{sinc}

\begin{document}
	
	\preprint{}
	
	\title{Photon Intensity Profiles for Four-Wave Mixing through a Kerr Medium}
	
	\author{P. Moodley}
	\email{Preshin.Moodley@gmail.com}

	\author{S. \surname{Roux}}
	\affiliation{National Metrology Institute of South Africa}

\date{March 11, 2024}

\begin{abstract}%
We study four wave mixing in a dielectric medium and calculate the detection probability of signal and idler photons landing on a screen. The 
outgoing photons are theoretically well characterized and can be used as probes for experimental investigations. The intensity plots are 
presented which compare well with experimental results.
\end{abstract}

\maketitle

\section{Introduction}
The field of nonlinear optics has attracted more attention in part due to its versatility as source for new technological innovation and for the 
illuminating effect it has on investigating the foundations of quantum mechanics and reality. The discovery of entangled photons pairs 
originating from Spontaneous Parametric Down Conversion (SPDC) processes hastened developed of efficient sources of SPDC photons. This has lead to improved bounds on the Bell inequality 
\cite{bell1964einstein,clauser1969proposed,lo2016experimental}, SPDC photons have also been used by exploiting their entanglement for quantum 
cryptography \cite{marcikic2004distribution,sergienko1999quantum}, ghost imaging \cite{strekalov1995observation} and metrology \cite{d2001two}. 
\\
\\
Four-wave mixing provides an advantage over three-wave mixing as it can result in the generation of new optical frequencies that are not available in three-wave mixing \cite{Klyshko,menzel2013photonics}. This opens up new opportunities for quantum information processing and the study of nonlinear optical phenomena.
Four-wave mixing was first studied in 1962 by Armstrong et. al \cite{armstrong1962} where a general framework for processes third order in the electric field was introduced. This lead to a wide variety of phenomena which could be described by variations in frequencies, wave vector and polarization directions \cite{shen1976recent,Levenson1980,carreira1978comparison}. Four-wave mixing is useful in quantum metrology for precision measurement of physical quantities, such as time and frequency. In addition, it is also applied in optical communication systems for the generation of frequency combs and in spectroscopy for the investigation of molecular interactions.
Recent innovative applications of four-wave-mixing have been used to develop low cost microscopy \cite{wang2021simplified}, as a tool for manipulating orbital angular momentum states for use in both classical and quantum communication \cite{offer2019spiral}, as a source for squeezed state light \cite{Bondurant1984}, improved limits for comb lasers near the visible range which are useful for atomic clocks and metrology \cite{Cruz08,xue2017second}. \\
\\
The purpose of this article is to study the optical Kerr effect and the detection probability of the photons originating from the Kerr effect falling on a screen. The Kerr effect is a four wave mixing process in which the refractive properties of a nonlinear medium are modified by the presence of four electric fields at a point and thus influencing the electromagnetic fields that are passing through it. We calculate the detection probability of photons falling on a screen after passing through a dielectric medium which facilitates the phase matching. The calculated intensity profile for the non degenerate case is sharp enough that the outgoing photons can be used as probes for further experimental investigations. \\
\\
This article is organized as follows: in Sec. II we introduce the Hamiltonian for the Kerr process and compute the transition amplitude for the two photon to two  photon process, in Sec. III we extract out the SPDC wave function, in Sec. IV we implement the conservation of momentum and energy to get the phase mismatch and in Sec. V we compute the detection probability of a photon falling on a screen originating from the Kerr process and we present the intensity plots and discuss the results.
%
\section{Hamiltonian  for the Kerr Effect}
\label{sec:hamiltonian}
Consider the case of two incoming photons passing through a crystal of length $L$ centered at the origin, interact non-linearly via the Kerr effect and exit as two outgoing photons.

\begin{figure}[!h]
	\centering
	\includegraphics[width=0.5\linewidth]{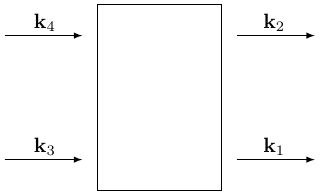}
	\caption{Incoming and outgoing states passing through the crystal.}
	\label{fig:photondiagram}
\end{figure}
The transition amplitude for the above process is given by
\begin{align}
	M =& \sum\limits_{p,q,r,s} {\int {V_{pqrs} \phi _p^{*\left( s \right)}\left( {{\boldsymbol k_1}} \right)\phi _q^{*\left( i \right)}\left( {{\boldsymbol k_2}} \right)\phi _r^{\left( {{p_1}} \right)}\left( {{\boldsymbol k_3}} \right)} }  \times \nonumber\\
	& \phi _s^{\left( {{p_2}} \right)}\left( {{\boldsymbol k_4}} \right)d{t_1}\frac{{{d^3}{k_1}}}{{{{\left( {2\pi } \right)}^3}{\omega _1}}}\frac{{{d^3}{k_2}}}{{{{\left( {2\pi } \right)}^3}{\omega _2}}}\frac{{{d^3}{k_3}}}{{{{\left( {2\pi } \right)}^3}{\omega _3}}}\frac{{{d^3}{k_4}}}{{{{\left( {2\pi } \right)}^3}{\omega _4}}} ,
\end{align}
defining the time independent part
\begin{align}
	{V_{pqrs}}: = \frac{i}{\hbar }\int\limits_{{t_0}}^t {\left\langle {{k_1},p;{k_2},q\left| {{H^{\left( i \right)}}\left( {{t_1}} \right)} \right|{k_3},r;{k_4},s} \right\rangle d{t_1}} .
\end{align}
This transition amplitude is interpreted as the Feynman rule for a vertex and is calculated once only as it is independent of the initial and final states that can vary. The Hamiltonian for the Kerr effect is given by
\begin{align}
	H:= \!\int\! {{\varepsilon _0}\chi _{zabc}^{\left( 3 \right)}{E_z}\left( {{t_1}, \boldsymbol x } \right){E_a}\left( {{t_1}, \boldsymbol x } \right){E_b}\left( {{t_1}, \boldsymbol x } \right){E_c}\left( {{t_1}, \boldsymbol x } \right){d^3}x} , \label{Hamiltonian}
\end{align}
where the $\chi_{zabc}$ is the nonlinear susceptibility and the volume integral is over the crystal.
The electric field decomposed in terms of the creation and annihilation operators
\begin{align}
	\boldsymbol E \left( { \boldsymbol x ,t} \right) = {\boldsymbol E ^{\left(  -  \right)}} + {\boldsymbol E ^{\left(  +  \right)}} , 
\end{align}
using the commutation rule with the definition of the ground states
\begin{align}
	\left[ {{a_s}\left( {{{\boldsymbol k }_1}} \right),a_r^\dag \left( {{{\boldsymbol k }_2}} \right)} \right] = {\left( {2\pi } \right)^3}{\omega _1}{\delta _{rs}}{\delta ^3}\left( {{{\boldsymbol k }_1} - {{\boldsymbol k }_2}} \right) ,  \\
	\left\langle {\boldsymbol k ,s} \right| : = \left\langle 0 \right|{a_s}\left( {\boldsymbol k } \right) \quad \left| {\boldsymbol k ,s} \right\rangle : = a_s^\dag \left( {\boldsymbol k } \right)\left| 0 \right\rangle ,
\end{align}
we get the action of the electric field operators on the vacuum. For the four fields at a point in \eqref{Hamiltonian}, only 6 of the 24 terms in the expression can connect the incoming and outgoing states. We can construct a dimensionless coupling parameter $g$ which absorbs all the constants by inserting suitable factors of $1=\frac{c}{c}$ into $V_{pqrs}$ giving the transition amplitude to be
\begin{align}
	{V_{pqrs}} = i{c^3}\int\limits {\int {} } {g}\frac{{{\omega _1}{\omega _2}{\omega _3}}}{{\omega _4^3}}{e^{i\left( {\Delta \omega {t_1} - \Delta \boldsymbol k  \cdot \boldsymbol x } \right)}}d{t_1}{d^3}x , \label{trans}
\end{align}
with the definitions
\begin{align}
	\Delta \omega &:= {\omega _1} + {\omega _2} - {\omega _3} - {\omega _4} \label{ConsLawsA}, \\
	\Delta \boldsymbol k &:= {\boldsymbol k _1} + {\boldsymbol k _2} - {\boldsymbol k _3} - {\boldsymbol k _4} . \label{ConsLawsB}
\end{align}

For the bright incoming beams of a laser, we can represent them as coherent states which contribute overall magnification factors to the single photon states which we can absorb into our effective coupling $g$. Assuming a narrow beam diameter which is small compared to the cross section of the crystal we can perform the integration over the orthogonal coordinates by extending the integration domain to infinity giving us delta functions over the transverse wave vector components. We also assume that the interaction time scale is extremely small in comparison to the passage of the beam through the crystal so that we can extend the time integral bounds to infinity giving us another delta function in the angular frequencies. The integration over the $z$ component of the crystal with length $L$ centered about the origin gives us a sinc function of the phase mismatch with the result
\begin{align}
	{V_{pqrs}} = {\left( {2\pi } \right)^3}{c^3}Lg\frac{{{\omega _1}{\omega _2}{\omega _3}}}{{\omega _4^3}}\delta \left( {\Delta \omega } \right)\delta \left( {\Delta {k_x}} \right)\delta \left( {\Delta {k_y}} \right)\times \nonumber\\
	\sinc\left( {\frac{1}{2}\Delta {k_z}L} \right) . \label{FeynVert2}
\end{align}
\section{State Extraction}
We now move to the preferred set of variables $\omega, k_x, k_y$ and with $k_z$ the dependent variable. Since we have chosen the beam to propagate along the $z$ axis, the system is considered to evolve along the $z$ axis and not as a function of time. Expressing the wave functions in terms of the transverse components 
\begin{align}
	\phi \left( {\boldsymbol k } \right)&: = c\psi \left( {{{\boldsymbol k }^ \bot },\omega } \right) , 
\end{align}
and changing the integration measure to the transverse components and the angular frequency such that the new integration measure is 
\begin{align}
	{k_z} &= \sqrt {\frac{{{n^2}{\omega ^2}}}{{{c^2}}} - {{\left| {{{\boldsymbol k }^ \bot }} \right|}^2}} \label{dispersion} ,\\
	\Rightarrow d{k_z} &= \frac{{{n^2}\omega d\omega }}{{{c^2}\sqrt {\frac{{{\omega ^2}}}{{{c^2}}} - {{\left| {{{\boldsymbol k }^ \bot }} \right|}^2}} }} = \frac{{{n^2}\omega d\omega }}{{{c^2}{k_z}}}. 
\end{align}
We can extract out the SPDC state by comparing the transition amplitude to 
\begin{align}
	M = \left\langle {\psi _p^{\left( s \right)},\psi _q^{\left( i \right)}|{\psi _{spdc}}} \right\rangle.
\end{align}
The incoming and going states are ideally of a single frequency, so we can assume the wave functions can be separated into transverse momentum and frequency parts which individually satisfy normalization conditions
\begin{align}
	\psi _r^{\left( {{p_1}} \right)}\left( {\boldsymbol k _3^ \bot } \right) &:= \sqrt {{k_{3z}}} {M_3}\left( {\boldsymbol k _3^ \bot } \right){H_3}\left( {{\omega _3}} \right) ,\label{SepWaveFnsA}\\
	\psi _s^{\left( {{p_2}} \right)}\left( {\boldsymbol k _4^ \bot } \right) &:= \sqrt {{k_{4z}}} {M_4}\left( {\boldsymbol k _4^ \bot } \right){H_4}\left( {{\omega _4}} \right) , \label{SepWaveFnsB}
\end{align}
We can combine the delta functions over the transverse momenta and do the integration over one of the incoming pumps variables. To enforce the single frequency states, we can define the angular frequency part of the wave function $H_a$ in terms of narrow Gaussians with small widths.
Assuming the momenta are almost collinear, we get
\begin{widetext}
	\begin{align}
		\left| {\psi {'_{spdc}}} \right\rangle =&  N\int {}  n_1^2n_2^2n_3^2n_4^2 \left| {\boldsymbol k _1^ \bot ,{\omega _1};\boldsymbol k _2^ \bot ,{\omega _2}} \right\rangle {M_3}\left( {\boldsymbol k _3^ \bot } \right){M_4}\left( {\boldsymbol k _4^ \bot } \right)\left( {\frac{{{\omega _3}}}{{\omega _4^7}}} \right)\sinc\left( {\frac{1}{2}\Delta {k_z}L} \right) \times \nonumber\\
		& \exp \left[ { - {{\left( {\frac{{{\omega _1} - {\omega _s}}}{{\delta {\omega _f}}}} \right)}^2} - {{\left( {\frac{{{\omega _2} - {\omega _i}}}{{\delta {\omega _f}}}} \right)}^2} - {{\left( {\frac{{{\omega _3} - {\omega _{{p_1}}}}}{{\delta \omega }}} \right)}^2} - {{\left( {\frac{{{\omega _4} - {\omega _{{p_2}}}}}{{\delta \omega }}} \right)}^2}} \right]
		d{\omega _1}d{\omega _2}d{\omega _3}\frac{{{d^2}k_1^ \bot }}{{{{\left( {2\pi } \right)}^3}}}\frac{{{d^2}k_2^ \bot }}{{{{\left( {2\pi } \right)}^3}}}\frac{{{d^2}k_3^ \bot }}{{{{\left( {2\pi } \right)}^3}}} , \label{psiprime}
	\end{align}
\end{widetext}
where $N$ is normalization constant, $\delta {\omega}$ and $\delta {\omega_f}$ are the band widths of the line filters. The Gaussians fall rapidly to zero outside their central frequencies. Computing the iterated integral over the $\omega_i$ we get approximately $\delta\omega^2_f \delta\omega$ with numerical factors which we absorb into the normalization constant. 
This allows the combined basis state to be separated into two independent states. The SPDC wave function can then be obtained by finding the components of the SPDC state in the basis $\left| {k_1^ \bot } \right\rangle \left| {k_2^\bot } \right\rangle$. The incoming pumps are approximated as Gaussian beams with the momentum wave functions
\begin{align}
	{M_3}\left( {{\boldsymbol k^\bot_3}} \right)&:= \sqrt {2\pi } {w_3}{e^{ - w_3^2{{\left| {{\boldsymbol k'^\bot_3-\boldsymbol q_3}} \right|}^2}}} ,\label{MomWaveFnA}\\
	{M_4}\left( {{\boldsymbol k^\bot_4}} \right)&:= \sqrt {2\pi } {w_4}{e^{ - w_4^2{{\left| {{\boldsymbol k'^\bot_4-\boldsymbol q_4}} \right|}^2}}} , \label{MomWaveFnB}
\end{align}
where $\boldsymbol k'^\bot_3=\boldsymbol k^\bot_3+\boldsymbol q_3$, $\boldsymbol k'^\bot_4=\boldsymbol k^\bot_4+\boldsymbol q_4$ are the incoming momentum wave vectors, $w_3,w_4$ are the widths and $\boldsymbol q_3,\boldsymbol q_4$ are shifts in the momentum of the incoming beams and allows to consider the case of non-collinear incoming beams. The SPDC wave function is given by
\begin{align}
	\psi {'_{spdc}} = 2\pi N{w_3}{w_4}\int {} {e^{ - w_3^2{{\left| {\boldsymbol k'^\bot_3 -\boldsymbol q_3} \right|}^2} - w_4^2{{\left| {\boldsymbol k'^\bot_4 -\boldsymbol q_4} \right|}^2}}} \times \nonumber\\ 
	\sinc\left( {\frac{1}{2}\Delta {k_z}L} \right)\frac{{{d^2}k_3^ \bot }}{{{{\left( {2\pi } \right)}^2}}} , \label{SPDCWaveFn}
\end{align}
where $\boldsymbol k^\bot_4$ is defined by setting the transverse components of \eqref{ConsLawsB} to zero. 
%
In the experiment we will pass the outgoing beams through a lens before it falls onto a detector screen. We will Fourier transform the SPDC wave function to pass from the transverse momenta to transverse spacial coordinates then pass the wave function through a 2f lens modeled as a Fourier transform finally yielding the wave function on the back focal plane
\begin{align}
	{\psi _{2\gamma}} \! \left( {\boldsymbol r _1^ \bot ,\boldsymbol r _2^ \bot } \right) \approx   \frac{1}{{{\lambda _1}{\lambda _2}{f^2}}}\psi {'_{spdc}} \! \left( {\boldsymbol k _1^ \bot ,\boldsymbol k _2^ \bot } \right) \!   \label{FourierTrans}
\end{align}
We evaluate the integral at 
\begin{align}
	\boldsymbol k _1^ \bot = \dfrac{{2\pi }}{{{\lambda _1}f}}\boldsymbol r _1^{\bot} \qquad\text{and}\qquad
	\boldsymbol k _2^ \bot = \dfrac{{2\pi }}{{{\lambda _2}f}}\boldsymbol r _2^{\bot}
\end{align}
where $\boldsymbol r _1^{\bot},\boldsymbol r _2^{\bot}$ are the radial distances on the screen and $f$ is the focal length of the lens. 
\section{Detection Probability}
We now have all the pieces to obtain the SPDC wave function at the back focal plane of the lens, substituting the phase mismatch from \eqref{PhaseMismatch} into \eqref{SPDCWaveFn} and \eqref{FourierTrans} using the integral representation of the $\sinc x$ function
\begin{align}
	\sinc x = \frac{1}{2}\int\limits_{ - 1}^1 {{e^{itx}}dt} ,
\end{align}
we can combine the arguments of the exponential and complete the square in $\boldsymbol k_3^\bot$ to perform the Gaussian integral over $\boldsymbol k_3^\bot$ to obtain
\begin{align}
	{\psi _{2\gamma }} &= \frac{N}{{w_4^2}}\int\limits_{ - 1}^1 {\frac{1}{{\delta \left( t \right)}}} {e^{\frac{1}{2}iL\Theta t + Q\left( t \right)}}dt ,
\end{align}
and with the definitions
\begin{align}
	{w_1}&: = \frac{{f{\lambda _1}}}{{2\pi {w_4}}} 		\qquad \qquad \qquad  \qquad \quad \!\! {w_2} := \frac{{f{\lambda _2}}}{{2\pi {w_3}}} ,\\
	s&: = {\left( {\frac{{{w_3}}}{{{w_4}}}} \right)^2} \qquad \qquad \qquad \quad \quad 	{z_R} := \frac{{\pi w_4^2}}{{{\lambda _4}}}  ,\\
	a&: = \sec {\theta _4} + \frac{{{\lambda _3}\sec {\theta _3}}}{{{\lambda _4}}} \qquad \qquad \,\,\, {\tau} := \frac{L}{{8{z_R}}} ,\\
	\Theta &: = \frac{{\pi {{\sin }^2}{\theta _1}}}{{{\lambda _1}\cos {\theta _1}}} + \frac{{\pi {{\sin }^2}{\theta _2}}}{{{\lambda _2}\cos {\theta _2}}} - \frac{{\pi {{\sin }^2}{\theta _3}}}{{{\lambda _3}\cos {\theta _3}}} - \frac{{\pi {{\sin }^2}{\theta _4}}}{{{\lambda _4}\cos {\theta _4}}}.
\end{align}
We have introduced a small dimensionless parameter $\tau \ll 1$ which is a ratio of the crystal length $L$ to the Rayleigh range $z_R$ which serves as an expansion parameter. We also introduce the auxiliary functions to keep the expressions compact
\begin{align}
	\alpha:=&1-it\tau\left(\sec\theta_4 - \frac{\lambda_1}{\lambda_4}\sec\theta_1\right), \\
	\beta :=&1-it\tau\left(\sec\theta_4 - \frac{\lambda_2}{\lambda_4}\sec\theta_2\right) ,\\
	\delta:=&1 + s - it\tau a, \qquad \gamma:=1 - it\tau \sec\theta_4,
\end{align}%
with the coefficient functions
\begin{align}
	A:=\alpha-\frac{\gamma^2}{\delta} \quad B:=\beta-\frac{\gamma^2}{\delta} \quad C:=\gamma-\frac{\gamma^2}{\delta}.
\end{align}
We can now express the argument of the exponential in the form
\begin{align}
	\!\!\!\!Q\left( t \right): =\!\! - A{\left| {\frac{{{\boldsymbol r_1}}}{{{w_1}}}} \right|^2} \!\!- B{\left| {\frac{{{\boldsymbol r_2}}}{{{w_2}}}} \right|^2} \!\!- 2C\left( {\frac{{{\boldsymbol r_1}}}{{{w_1}}}} \right) \cdot \left( {\frac{{{\boldsymbol r_2}}}{{{w_2}}}} \right). \label{Q}
\end{align}
We will keep only terms of order $\tau$ and ignore higher order terms for this calculation since we are using the thin crystal approximation. We can expand out the coefficient functions to obtain
\begin{align}
	{c_1}&: = \frac{s}{{1 + s}} \qquad \qquad \quad {c_2}: = \frac{{\sec {\theta _4}\left( {1-{s^2}} \right)}}{{{{\left( {1 + s} \right)}^2}}},\\
	{c_3}&: = {c_2} + \frac{{{\lambda _1}}}{{{\lambda _4}}}\sec {\theta _1}  \quad \, \, \, {c_4}: = {c_2} + \frac{{{\lambda _2}}}{{{\lambda _4}}}\sec {\theta _2},
\end{align}
with the order $\tau$ expansion for the coefficient functions 
\begin{align}
	A&: = {c_1} + it\tau {c_3} \quad B: = {c_1} + it\tau {c_4} \quad C: = {c_1} + it\tau {c_2}.
\end{align}
The probability distribution of a photon falling on the screen is given by
\begin{align}
	P\left( {\boldsymbol r _1^ \bot} \right) &:= \int {{{\left| {{\psi _{2\gamma }}\left( {{\boldsymbol r ^ \bot_1},{\boldsymbol r ^ \bot_2}} \right)} \right|}^2}{d^2}{\boldsymbol r ^ \bot_2}}. \label{psisq}
\end{align}
The integral over $\boldsymbol r^\bot_2$ can be done by completing the square of \eqref{Q} and integrating the resulting Gaussian, leaving the parameter integral over the $t_i$'s. Defining the functions below which are the sums of the $A$'s, $B$'s and $C$'s with their complex conjugates which come from the wave function squared in \eqref{psisq}
\begin{align}
	U\left( {{t_1},{t_2}} \right): = 2{c_1} + i\tau {c_3}\left( {{t_1} - {t_2}} \right),\\
	V\left( {{t_1},{t_2}} \right): = 2{c_1} + i\tau {c_4}\left( {{t_1} - {t_2}} \right),\\
	W\left( {{t_1},{t_2}} \right): = 2{c_1} + i\tau {c_2}\left( {{t_1} - {t_2}} \right),
\end{align}
giving us the probability distribution
\begin{widetext}
	\begin{align}
		\!\!P =& \frac{{\pi w_2^2}}{{w_4^4}}N\int\limits_{ - 1}^1 {\int\limits_{ - 1}^1 {} } {\frac{1}{{\delta \left( {{t_1}} \right){\delta ^*}\left( {{t_2}} \right)V}}} \exp \left[ {\frac{1}{2}iL\Theta \left( {{t_1} - {t_2}} \right)- \left( {U - \frac{{{W^2}}}{V}} \right){{\left| {\frac{{{\boldsymbol r_1}}}{{{w_1}}}} \right|}^2} } \right]d{t_1}d{t_2} \label{parameterint}.
	\end{align}
\end{widetext}
The integrand of the parameter integral \eqref{parameterint} is not a rational function, and we have no hope of finding an elementary primitive. We can Taylor expand the coefficients of the integrand and the argument of the exponential keeping only terms up to first order in the expansion parameter $\tau$ then integrate over the $t_1$ and $t_2$. With the definitions
\begin{align}
	{\tilde{R}^2}&: = {\left| {\frac{{\boldsymbol r_1^ \bot }}{{{w_1}}}} \right|^2},\label{defRsq} \qquad \qquad  \\
	\tilde{R}_0^2&: =\frac{1}{4}L\Theta \cos {\theta _4},\label{defR0sq}
\end{align}
we obtain the probability distribution 
\begin{align}
	P = \frac{{2w_2^2{N^2}}}{{w_3^2\left( {w_3^2 + w_4^2} \right)}}{{\mathop{\rm sinc}\nolimits} ^2}\left[ {2\sec {\theta _4}\tau \left( {{\tilde{R}^2} - \tilde{R}_0^2} \right)} \right]. \label{FinalP}
\end{align}
%
\section{Results}
Using a crystal length of $L=2$ mm and incoming beam with wavelengths $\lambda_3=\lambda_4=633$ nm and beam widths $w_2=w_3=w_4=1$ mm we can plot the probability distribution. Since \eqref{FinalP} is a function of incoming beam angle and the radial distance on the screen, we obtain concentric rings at the peaks of the probability distribution. Let $x$ be the scaled radial distance to the width $w_1$ from the origin on the screen. 
%
\begin{figure}[!h]
	\includegraphics[clip,width=\columnwidth]{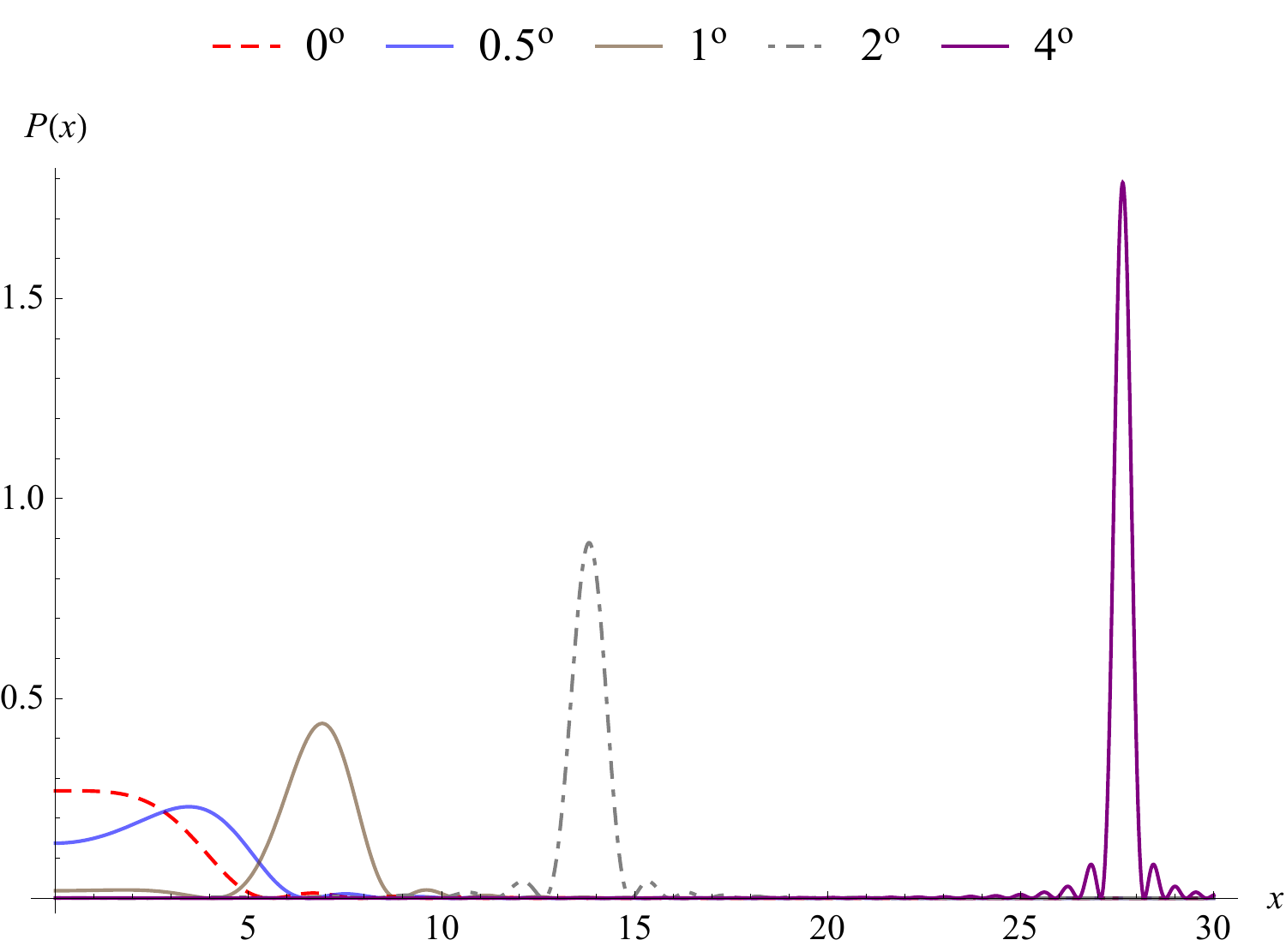}%
	\caption{Probability distribution for collinear incoming beams and varying outgoing beam for $\theta_1\in \{0^\circ, 0.5^\circ, 1^\circ, 2^\circ, 4^\circ \}$}\label{fig:prob1a}
\end{figure}
For the case of collinear incoming beams and varying outgoing beam we have the distributions as shown in Fig.~\ref{fig:prob1a}. We see the broad featureless Gaussian blob in the case when $\theta_1 = 0^\circ$ centered about the origin. For the 
$\theta_1 \neq 0^\circ$ cases we see some varying peaks with the amplitude decaying quickly outside the central peak. The location of the peaks moves further away from the origin with increasing angle and the widths of the central peaks get smaller with increasing outgoing angle.

\begin{figure}[!h]
	\includegraphics[clip,width=\columnwidth]{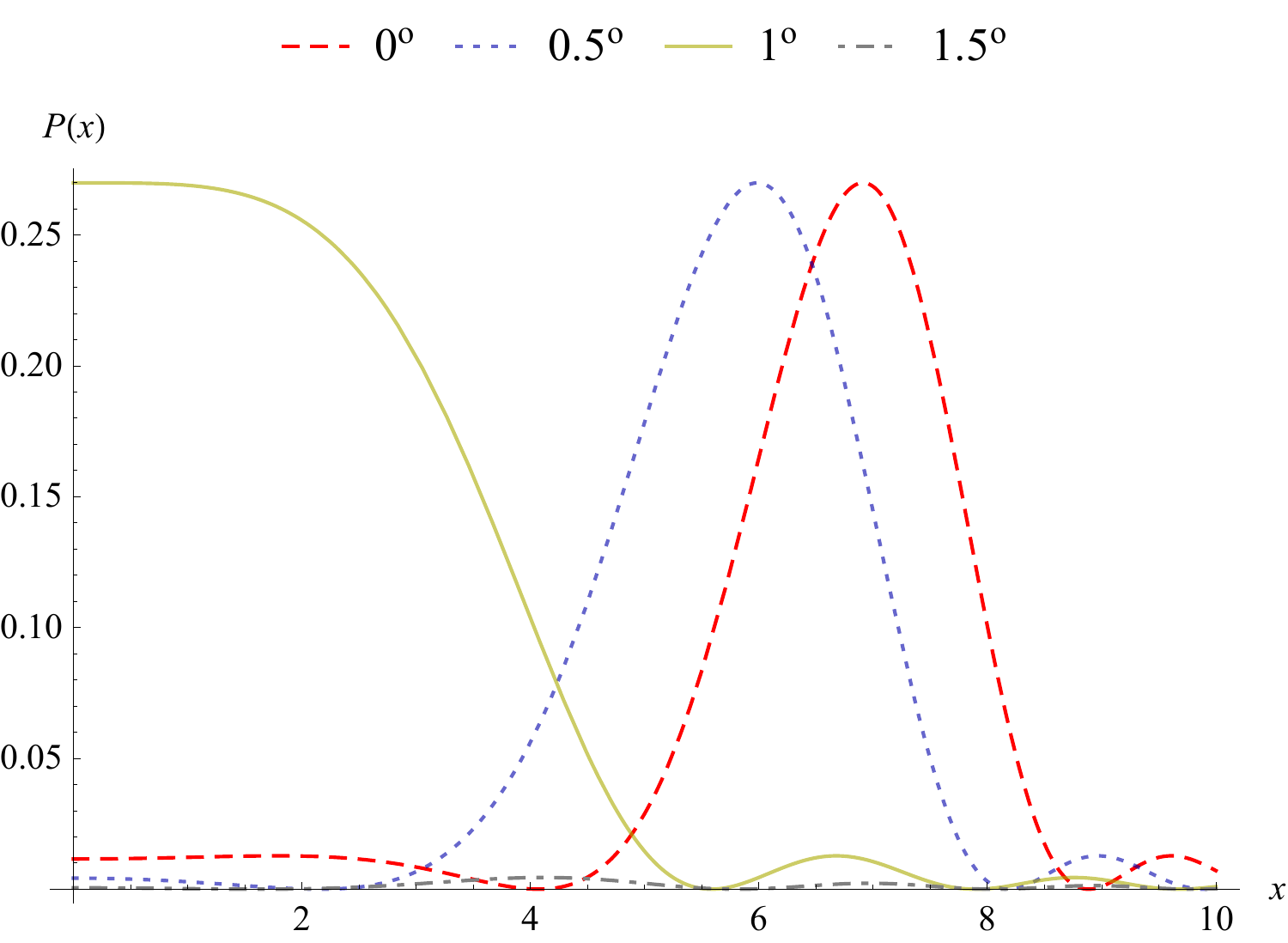}
	\caption{Probability distribution for varying incoming beams $\theta_4 \in \{0^\circ, 0.5^\circ, 1^\circ, 1.5^\circ \}$ and fixed outgoing beam for $\theta_1 = 1^\circ$}\label{fig:prob2a}
\end{figure}
For the non collinear case of varying the incoming beam angle while keeping the outgoing beam angle fixed we have the distributions  as shown in Fig.~\ref{fig:prob2a}. We see that as the incoming angle is increased from $\theta_4 =0^\circ$ the central peaks move closer to the origin and the widths of the central peaks get wider. 

\section{Conclusion}
We obtained concentric rings at the peaks of the probability distribution. Experimental results for the rings can be seen in 
\cite{Maker,armstrong1962}. The photons in the outgoing beams lie on two concentric cones, which in this case lie on top of each other since we 
are dealing with the degenerate case for the outgoing angles. The pair of photons fall on the screen forming rings at opposite points on the 
ring. The peak intensity corresponds to the location where the critical phase matching condition was almost satisfied with a sharp drop off 
beyond this point. \\
\\
We have presented the calculation to obtain the detection probability for a photon falling on a screen originating from a four-wave mixing process in a Kerr medium. We made use of the thin crystal approximation to linearize the integrand of the probability distribution which was then plotted as a function of radius on the screen. This provides a useful tool for the study and understanding of this nonlinear optical process and allows for a deeper analysis of the underlying physics which could be useful for the optimization of various applications, such as in quantum metrology and optical communication systems.
%
%
%
\begin{acknowledgments}
	This work was supported in part by funding 127523 from the National Research Foundation of South Africa. P.M. thanks Bertus Jordaan and Sefako Mofokeng for discussions on quantum optics and experimental setups in optics.
\end{acknowledgments}
\appendix
\section{Phase Mismatch}
The phase mismatch is encoded in the sinc function of \eqref{SPDCWaveFn}. We can express the mismatch in terms of transverse components of three of the wave vectors, by setting the transverse component of \eqref{ConsLawsB} to zero. The phase mismatch can be expressed using the dispersion relation \eqref{dispersion}. For the case of the two outgoing beams exiting the crystal at some nonzero angles, we associate an $\varepsilon$ with the transverse component of each beam to keep track of the part of the expression that is small under the paraxial approximation. The $z$ component of the $m^{th}$ wave vector is give by
\begin{align}
	{k_{mz}} = \sqrt {{{\left( {\frac{{2\pi }}{{{\lambda _m}}}} \right)}^2} - \boldsymbol k_m^{ \bot 2}} 
\end{align}
We will tag the individual wave vectors using an $\varepsilon$ in the form of 
\begin{align}
	\boldsymbol k_m^{ \bot 2} \to {\left( {\frac{{2\pi }}{{{\lambda _m}}}} \right)^2}{\sin ^2}{\theta _m} + \varepsilon \left[ {\boldsymbol k_m^{ \bot 2} - {{\left( {\frac{{2\pi }}{{{\lambda _m}}}} \right)}^2}{{\sin }^2}{\theta _m}} \right] \label{tagging}
\end{align}
where the $\theta_m$ refer to the angles the outgoing beams make with the beam axis. We can approximate the $z$ component as
\begin{align}
	{k_{mz}} &= \sqrt {{{\left( {\frac{{2\pi }}{{{\lambda _m}}}} \right)}^2} - \boldsymbol k_m^{ \bot 2}} \nonumber\\
	&\simeq \frac{{2\pi }}{{{\lambda _m}}}\cos {\theta _m} - \frac{\varepsilon }{2}\left[ {\frac{{{\lambda _m}}}{{2\pi }}\frac{{\boldsymbol k_m^{ \bot 2}}}{{\cos {\theta _m}}} - \frac{{2\pi {{\sin }^2}{\theta _m}}}{{{\lambda _m}\cos {\theta _m}}}} \right]
\end{align}
We can now calculate the phase mismatch as
\begin{widetext}
	\begin{align}
		\Delta {k_z} =& {k_{1z}} + {k_{2z}} - {k_{3z}} - {k_{4z}} \nonumber\\
		=& 2\pi \left( {\frac{{\cos {\theta _1}}}{{{\lambda _1}}} + \frac{{\cos {\theta _2}}}{{{\lambda _2}}} - \frac{{\cos {\theta _3}}}{{{\lambda _3}}} - \frac{{\cos {\theta _4}}}{{{\lambda _4}}}} \right) + \nonumber\\
		& - \frac{\varepsilon }{2}\left[ {\frac{{{\lambda _1}}}{{2\pi }}\frac{{ \boldsymbol k_1^{ \bot 2}}}{{\cos {\theta _1}}} + \frac{{{\lambda _2}}}{{2\pi }}\frac{{ \boldsymbol k_2^{ \bot 2}}}{{\cos {\theta _2}}} - \frac{{{\lambda _3}}}{{2\pi }}\frac{{ \boldsymbol k_3^{ \bot 2}}}{{\cos {\theta _3}}} - \frac{{{\lambda _4}}}{{2\pi }}\frac{{ \boldsymbol k_4^{ \bot 2}}}{{\cos {\theta _4}}} - \frac{{2\pi {{\sin }^2}{\theta _1}}}{{{\lambda _1}\cos {\theta _1}}} - \frac{{2\pi {{\sin }^2}{\theta _2}}}{{{\lambda _2}\cos {\theta _2}}} + \frac{{2\pi {{\sin }^2}{\theta _3}}}{{{\lambda _3}\cos {\theta _3}}} + \frac{{2\pi {{\sin }^2}{\theta _4}}}{{{\lambda _4}\cos {\theta _4}}}} \right]
	\end{align}
For critical phase matching we set $\varepsilon=1$ and 
	\begin{align}
		{\frac{{\cos {\theta _1}}}{{{\lambda _1}}} + \frac{{\cos {\theta _2}}}{{{\lambda _2}}} - \frac{{\cos {\theta _3}}}{{{\lambda _3}}} - \frac{{\cos {\theta _4}}}{{{\lambda _4}}}} = 0 \label{consE}
	\end{align}
giving us
	\begin{align}
		\Delta k_z^{crit} = \frac{{{\lambda _3}}}{{4\pi }}\frac{{ \boldsymbol k_3^{ \bot 2}}}{{\cos {\theta _3}}} + \frac{{{\lambda _4}}}{{4\pi }}\frac{{ \boldsymbol k_4^{ \bot 2}}}{{\cos {\theta _4}}} - \frac{{{\lambda _1}}}{{4\pi }}\frac{{ \boldsymbol k_1^{ \bot 2}}}{{\cos {\theta _1}}} - \frac{{{\lambda _2}}}{{4\pi }}\frac{{ \boldsymbol k_2^{ \bot 2}}}{{\cos {\theta _2}}} + \frac{{\pi {{\sin }^2}{\theta _1}}}{{{\lambda _1}\cos {\theta _1}}} + \frac{{\pi {{\sin }^2}{\theta _2}}}{{{\lambda _2}\cos {\theta _2}}} - \frac{{\pi {{\sin }^2}{\theta _3}}}{{{\lambda _3}\cos {\theta _3}}} - \frac{{\pi {{\sin }^2}{\theta _4}}}{{{\lambda _4}\cos {\theta _4}}}
	\end{align}

We now use the conservation of momentum
	\begin{align}
		\boldsymbol k_1^ \bot  + \boldsymbol k_2^ \bot  = \boldsymbol k_3^ \bot  + \boldsymbol k_4^ \bot 
	\end{align}
	to solve for $\boldsymbol k_4^ \bot$ and substitute it into the phase match equation giving us
	\begin{align}
		\Delta k_z^{crit} =& \left( {\frac{{{\lambda _4}}}{{\cos {\theta _4}}} - \frac{{{\lambda _1}}}{{\cos {\theta _1}}}} \right)\frac{{ \boldsymbol k_1^{ \bot 2}}}{{4\pi }} + \left( {\frac{{{\lambda _4}}}{{\cos {\theta _4}}} - \frac{{{\lambda _2}}}{{\cos {\theta _2}}}} \right)\frac{{ \boldsymbol k_2^{ \bot 2}}}{{4\pi }} + \left( {\frac{{{\lambda _3}}}{{\cos {\theta _3}}} + \frac{{{\lambda _4}}}{{\cos {\theta _4}}}} \right)\frac{{ \boldsymbol k_3^{ \bot 2}}}{{4\pi }} + \nonumber\\
		&+ \frac{1}{{4\pi }}\frac{{2{\lambda _4}}}{{\cos {\theta _4}}}\left( { \boldsymbol k_1^ \bot  \boldsymbol k_2^ \bot  -  \boldsymbol k_1^ \bot  \boldsymbol k_3^ \bot  -  \boldsymbol k_2^ \bot  \boldsymbol k_3^ \bot } \right) + \frac{{\pi {{\sin }^2}{\theta _1}}}{{{\lambda _1}\cos {\theta _1}}} + \frac{{\pi {{\sin }^2}{\theta _2}}}{{{\lambda _2}\cos {\theta _2}}} - \frac{{\pi {{\sin }^2}{\theta _3}}}{{{\lambda _3}\cos {\theta _3}}} - \frac{{\pi {{\sin }^2}{\theta _4}}}{{{\lambda _4}\cos {\theta _4}}} \label{PhaseMismatch}
	\end{align}
Where $\lambda_4$ is defined through \eqref{consE}.
\end{widetext}

\bibliographystyle{apsrev-title}

\bibliography{mybib}

\end{document}